# Ionic liquid/PDMS coated nanostructured ZnO surfaces for superior antibacterial activity

**Anushika Rohilla [a], Ajit Seth [a†], Mayank Bharadwaj [a], Rumal Singh [a], Sajal K Ghosh [a*]**

[a] *Department of Physics, Shiv Nadar Institution of Eminence, NH-91, Tehsil Dadri, Gautam Buddha Nagar, UP-201314, India*

*†Present Address: Institut für Experimentelle und Angewandte Physik, Christian-Albrechts-Universität zu Kiel, Kiel 24098, Germany*

## Abstract

The development of nanostructured antibacterial surfaces has gained significant attention due to the urgent need to combat bacterial colonization and biofouling across food, healthcare, and marine sectors. Among functional metal oxides, zinc oxide (ZnO) nanostructures are particularly promising owing to their rich morphology, favourable physicochemical attributes, and intrinsic bactericidal properties. Here, we utilize a green, eco-friendly 'Hot Water Treatment' (HWT) method to fabricate dense ZnO nanorod arrays on zinc substrates. To overcome the practical limitations and challenges of conventional antibacterial coatings, such as poor operational stability, cytotoxicity, and the risk of antimicrobial resistance, we developed a durable hybrid coating composed of an imidazolium-based ionic liquid stabilized with polydimethylsiloxane. The resulting hybrid-coated nanostructured surfaces exhibit superior antibacterial efficacy against *Escherichia coli* through a combination of topography-induced mechanical strain and membrane disruption.

***Corresponding author:*** sajal.ghosh@snu.edu.in

## Table of content figure

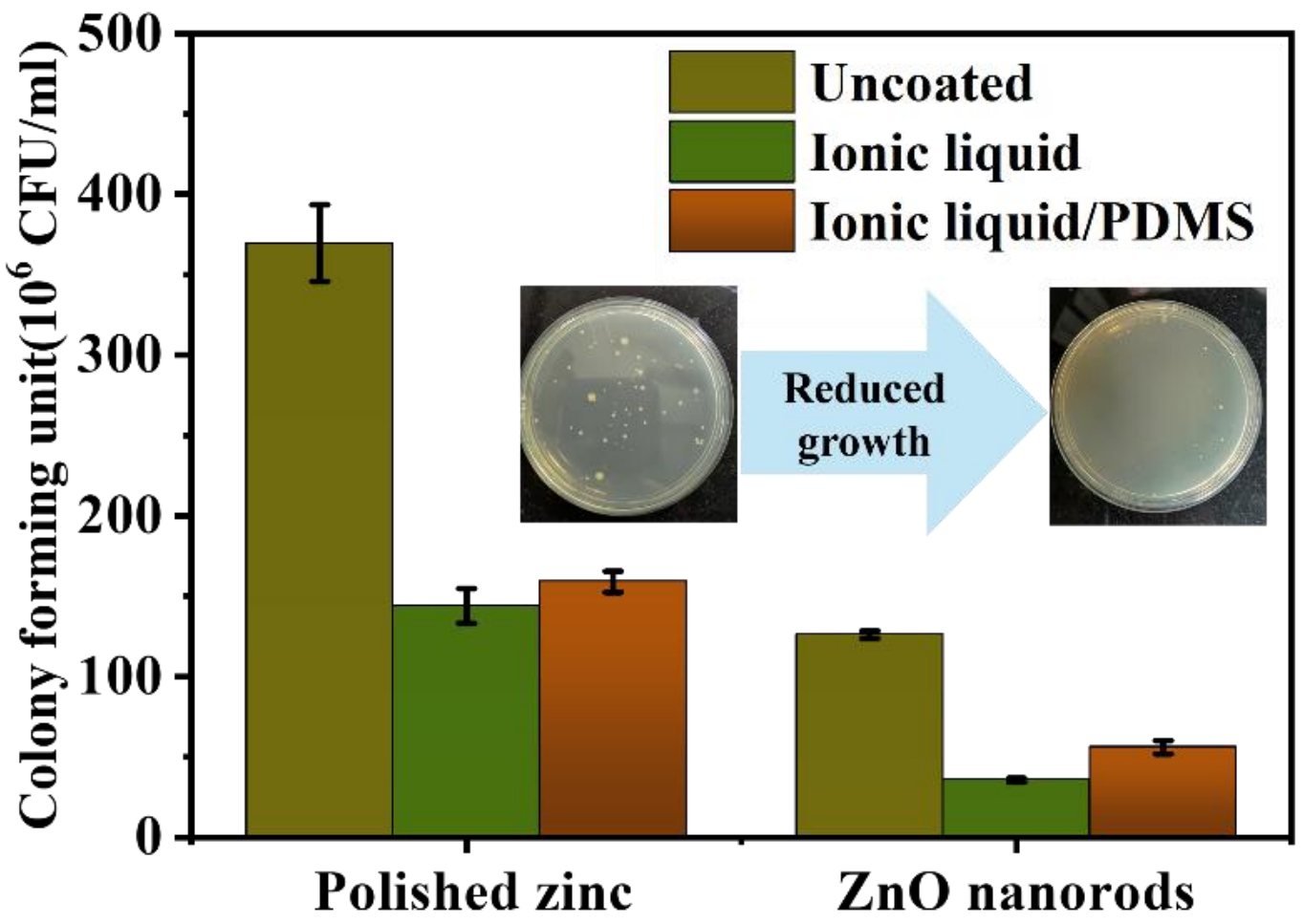

# 1. Introduction

The persistent challenges of bacterial contamination on surfaces pose a serious threat to public health, particularly in biomedical appliances where bacterial colonization can rapidly form biofilms.[1,2] Biofilms also contribute to industrial biofouling, corrosion, product spoilage, and the dissemination of pathogenic microorganisms, resulting in significant economic and societal burdens.[3]

A robust cell wall, which consists of a lipid layer reinforced by a peptidoglycan network, provides enhanced mechanical rigidity and structural support to bacterial cell.[4] This cell-wall architecture helps bacteria to withstand turgor pressures, preserving membrane integrity and cellular viability.[5] Nature has also evolved strategies to overcome bacterial infections, such as the array of nano-protrusions on the surfaces of cicadas and dragonflies, which have intrinsic bactericidal properties.[6] Unlike conventional antimicrobial systems that depend on the release of biocidal molecules, these surfaces physically interact with bacteria, damaging the cell wall and leading to the eventual cell death. This has been encouraging researchers across the fields to develop nature inspired antimicrobial and antibacterial surfaces for various applications.

Pogodin et al. proposed a model in which bacterial adhesion to nanopatterned surfaces generates membrane stretching sufficient to induce mechanical failure, identifying cell rigidity as a critical determinant of susceptibility.[5] Subsequent elastic mechanical models predicted that Gram-positive bacteria exhibit enhanced resistance, whereas Gram-negative cells are generally more vulnerable due to their lower mechanical stiffness.[7] Also, Xue *et al.* presented the theoretical model for the bactericidal nature of nanopatterned surfaces.[8] Later, thermodynamic approaches further demonstrated that bactericidal activity is governed by the balance between adhesion energy and membrane deformation energy, highlighting the importance of geometric factors including radius, shape and interspacing of nanostructures, in controlling antibacterial performance.[9]

Advances in biophysics have revealed that bacterial adhesion to surfaces is governed by a complex interplay of interfacial forces and surface energy, manifested as a difference in their wettability.[10] Therefore, along with the surface's physical structure, its chemical nature plays an important role. Nanopatterned surfaces not only physically rupture bacterial membranes but can also trigger oxidative stress pathways, inhibit cellular division, and alter bacterial physiology, indicating that mechanical and biochemical processes operate synergistically during bacterial killing.[11] Collectively, these findings have transformed antibacterial surface design into a multidisciplinary framework considering physics and chemistry of a surface.

To play with the chemistry of the surface, various materials have been utilized to coat a nanostructured surface with a very thin layer. Sarkar et al. developed a bilayer sol–gel coating consisting of a base layer containing ZnO nanoparticles and a flower-like ZnO top layer modified with 1H,1H,2H,2H-perfluorooctyl triethoxysilane (PFOTES), prepared by brush coating to obtain antimicrobial and superhydrophobic properties.[12] ZnO nanostructured surfaces hydrophobized with either stearic acid or fluoroalkylsilane (FAS) exhibit self-cleaning, bacteria-repellent properties.[13] These antibacterial coatings based on biocidal agents can effectively reduce bacterial contamination, but their cytotoxicity and potential to act as environmental pollutants remain a major concern. Many of them show limitations in terms of long-term stability due to the leaching of active compounds.[11] These limitations have motivated the development of sustainable antibacterial surfaces that combine durable surface chemistry with bactericidal nanotopography.

As antimicrobial metal, zinc oxide (ZnO) is attractive because of its photocatalytic activity and ability to produce reactive oxygen species (ROS) which disrupt bacterial membranes and causing cell death [14]. Importantly, ZnO nanostructures can be synthesized through green, eco-friendly routes that avoid toxic precursors and harsh conditions, aligning with the principles of sustainable nanotechnology. By tailoring the surface energy and morphology of the ZnO surface, it is possible to modulate wettability and thereby enhance antibacterial performance. In this context, ionic liquids (ILs) have gained attention as versatile antibacterial agents due to their tunable physicochemical properties and ability to destabilize microbial membranes.[15] Our previous studies demonstrated that IL coatings impart antibacterial activity to surfaces; however, the coatings were unstable, limiting their long-term effectiveness[15]. To address these challenges, the present work combines green-synthesized ZnO nanostructures with a polymer-stabilized ionic liquid coating to create a multifunctional antibacterial interface. By integrating nanoscale topographical features that induce bacterial deformation with the membrane-disruptive properties of ionic liquids, this strategy seeks to exploit both physical and chemical antibacterial mechanisms simultaneously. The resulting hybrid platform provides an opportunity to investigate how surface morphology, wettability, and interfacial chemistry collectively influence bacterial viability, while offering a robust and sustainable route towards developing next-generation antibacterial surfaces.

# 2. Materials and methods

## 2.1 Materials

Commercially available pure zinc plates, measuring 100 (L) × 100 (W) × 1 (H) $mm^3$, were obtained from Special Metals (Mumbai, India). 1-decyl-3-methylimidazolium chloride (DMIMCl) and polydimethylsiloxane (PDMS) were purchased from Sigma-Aldrich. Luria-Bertani (LB) broth and LB agar medium for the bacterial culture study were obtained from Hi-Media (Mumbai, India). The spectroscopic grade methanol was obtained from Thermo Fisher Scientific (USA). All the chemicals were being used without further purification. Deionised (DI) water (Milli-Q) with a resistivity of~18 MΩ.cm and a pH of~7.0 was used for all purposes.

## 2.2 Synthesis of ZnO nanostructures

The substrates, measuring 15 × 10 $mm^2$, were cut from commercially available pure zinc plates and then flattened at a pressure of 100 bar using a hydraulic press. These substrates were then polished with abrasive papers from 800 to 3000 grit size to impart microscale roughness. The flat polished zinc substrates were ultrasonically cleaned with methanol and DI water, respectively each for 15 minutes, to remove any organic contaminants. A gentle flow of nitrogen gas was used to dry the substrates after cleaning. Nanostructures on ZnO substrate were synthesized by hot water treatment which is an eco-friendly and cost-effective approach. The details of this technique have been described elsewhere.[16,17] In brief, the cleaned metal substrates were placed in hot water. At the water-metal interface, chemical reactions release metal ions, which then migrate and redeposit into nanostructures of various forms, such as nanorods, nanowires, nanosheets, and nanoflowers.[17] In the present study, cleaned Zn substrates were kept in deionised water at 75 ºC for 5 hours, after which, the substrates were dried using $N_2$ gas. These substrates with nanostructures were then heated in a hot air oven (Ambinova Technologies, India) at 200 ºC for 2 hours before further characterisation. The surface morphology of these substrates was investigated using Field Emission Scanning Electron Microscopy (FESEM) (JEOL JSM-7610 PLUS), and images were obtained by operating it at an accelerating voltage of 15 kV.

## 2.3 Coating of substrates

All the samples were spin coated using a single-chain ionic liquid DMIMCl and a composite of DMIMCl and PDMS. The cleaned substrates were kept in a UV ozone system (PSD Pro Series, Novascan Technologies, USA) at 50 ºC for 30 minutes with 10 minutes of hold time. It makes a surface hydrophilic by removing all the organic impurities. A chloroform solution containing 0.3 wt% DMIMCl was prepared by vortexing for 5 minutes. 50 µl of this solution was pipetted onto the zinc and ZnO nanostructured substrate and then spin-coated at 4000 rpm for 1 minute, followed by annealing at 100°C for 30 min.[15] For coating with a mixture of DMIMCl and PDMS, their separate solutions were prepared by taking 0.3 wt% in chloroform. Then they were mixed to obtain a final solution containing 95 wt% DMIMCl and 5 wt% PDMS. Spin coating was performed on the substrates using 50 µl of this mix at 500 rpm for 30 seconds, then at 5000 rpm for 1 minute. The coated substrates were then annealed at 95 ° C for 12 h.[18]

**2.4 Contact angle and adhesion force**

An optical tensiometer from One Attention, Theta Flex (Biolin Scientific) was used to measure the contact angle of a sessile drop of water of volume ~2 µl. The measurements were taken at five different spots on each substrate to obtain an average value with a standard deviation. The adhesion force of water with the substrate was measured using a force tensiometer (Attention Sigma701, Biolin Scientific). The reported values here are the average of three independent measurements on different substrate locations.[19]

**2.5 Antibacterial activity**

A nutrient-rich medium for the bacteria culture is Luria-Bertani (LB) broth, which is made up of casein enzymic hydrolysate, yeast extract, sodium chloride, and agar. 40 g of this composition was dissolved in 1000 mL of DI water, which was autoclaved at 120 °C for 20 minutes to sterilise, and then cooled to 60 °C. It was then poured into a sterile petri dish. These petri dishes were wrapped in parafilm after solidification and kept at 4 °C for further use. The yeast extract provided vitamin B complexes for bacterial growth, while the enzymatic hydrolysate produced peptides and peptones. Sodium chloride maintains osmotic equilibrium in the medium and supports membrane transport. The primary culture was prepared by incubating a glycerol stock of Gram-negative bacterium *E. coli* DH5α [20] and growing it overnight at 37 °C in LB broth medium with orbital shaking at 200 rpm.[15,21]After the overnight growth, 1 mL volume of this culture was dissolved to fresh LB broth to attain a final optical density (OD600) of 0.5. A 50 µL drop of this solution was dispensed on the fabricated substrates and then incubated for 6 hours at 37 °C. Then, the cells were taken out from the surfaces, pipetted in an aqueous solution of 0.9% sodium chloride, and kept in a sterile Eppendorf tube. For the spot assay, 5 µL of the solution at different dilutions was spotted onto LB agar plates (Hi-Media, Mumbai) and incubated overnight at 37 °C. The number of colonies were determined by manual counting. To calculate colony-forming units (CFUs), a 100 µL solution of the specified dilution was spread uniformly onto a sterile LB agar plate. It was allowed to dry for a few minutes at room temperature and thereafter, stored at 37 °C for 12 hours. CFU per mL is calculated using the relation, $CFU\ (ml) = \frac{(number\ of\ colonies \times Dilution\ factor)}{Volume\ of\ culture\ spread\ on\ agar\ plate\ (mL)}$.

# 3. Results and Discussion

## 3.1 Morphology and wetting nature of surfaces

The physical morphology and chemical composition of surfaces play a vital role in wettability, leading to differences in their antibacterial activities. Therefore, it is important to vary these two factors to develop an effective antibacterial surface. Figure 1 (a, b) represents the unpolished Zn substrate, with an irregular pattern and defects. After polishing these surfaces, a smoother morphology is observed in Figures 1(c) and 1(d) with micro-scale scratches.[22] The ZnO nanorods were fabricated on these substrates after hot water treatment (HWT) (Figures 1 (e, f)). The images reveal a densely packed, uniform distribution of nanorods exhibiting the characteristics of the wurtzite crystal structure of ZnO.[23] The orientation of these nanorods suggests preferential crystal growth along the polar C-axis to minimize surface free energy.[24] The formation of nanorods increases the surface-to-volume ratio and decreases the solid fraction compared with a flat, polished zinc substrate. As ZnO is inherently hydrophilic and air is hydrophobic, the entrapment of air within the nanorods modulates its wettability. Therefore, the contact angle and wetting condition of a water droplet on a rough surface are determined by the solid fraction. Depending on this fraction, the surface can either wet homogeneously in the Wenzel state, in which water enters the grooves, resulting in high adhesion. On the other hand, the surface can be wet heterogeneously due to a higher air fraction, known as the Cassie-Baxter state, in which water droplets exhibit low adhesion.[25]

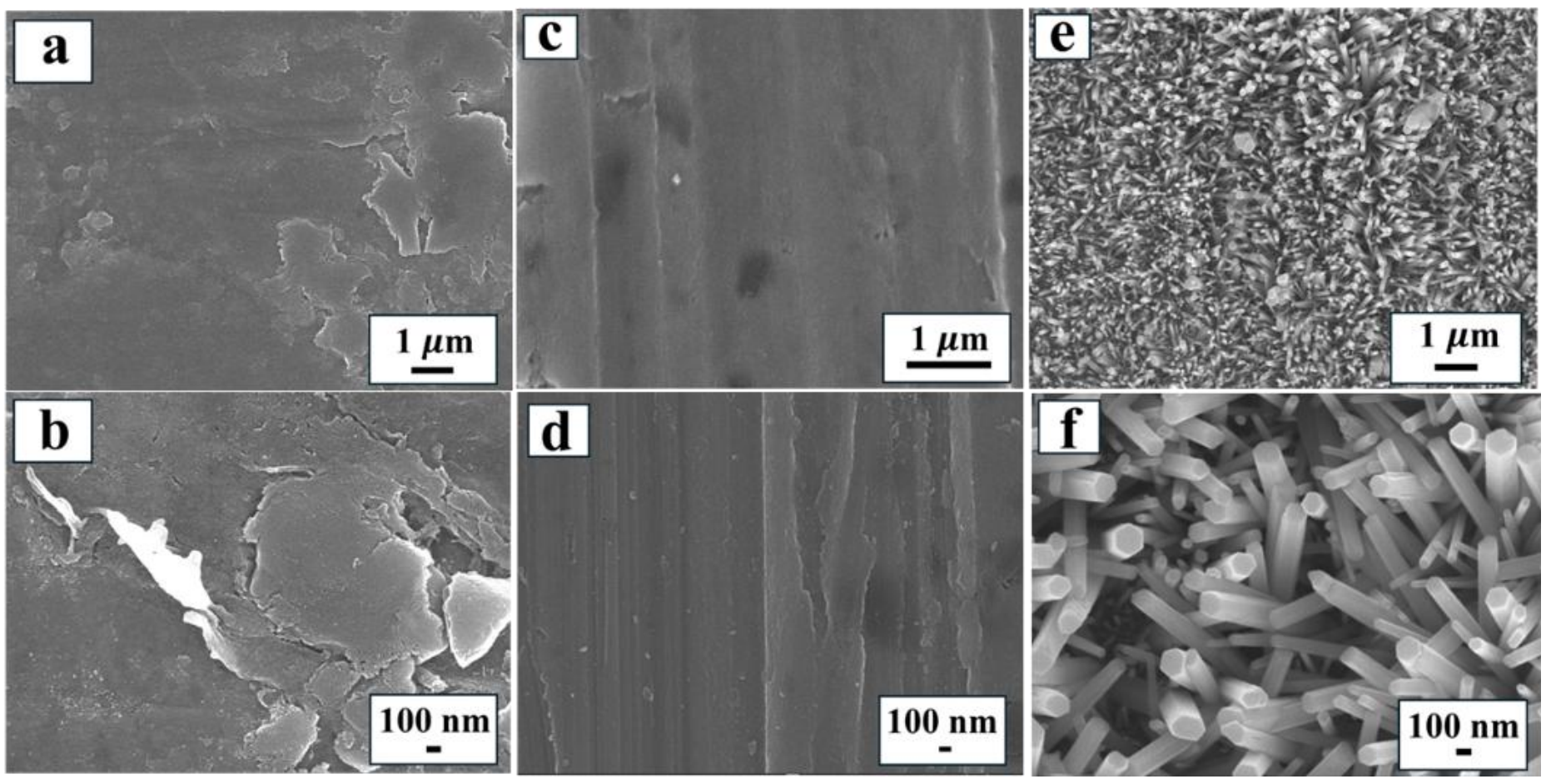

*Figure 1. Field emission scanning electron microscopy (FESEM) images of (a) unpolished zinc with zoomed image in (b), (c) polished zinc with zoomed image in (d) and (e) ZnO nanorods with zoomed images in (f).*

Surface wettability is a critical parameter when evaluating nanostructured surfaces for antibacterial applications, as it directly influences how nanostructures interact with bacterial cells. As the chemical nature of the surface also controls this interaction, the zinc and ZnO nanostructured surfaces were chemically modified using imidazolium-based ionic liquid (IL) DMIMCl and a composite of this IL with PDMS. The contact angle values of surfaces are shown in Figure 2(a). The surfaces are classified as hydrophilic with a contact angle less than 90°, while those exceeding 90° are hydrophobic.[26] The pristine zinc substrates are intrinsically hydrophilic, with a contact angle of 64° ± 4°, due to the pre-existing high-energy oxide layer. There is a significant transition in wettability upon the growth of nanorods, with the contact angle rising from 64° ± 4 ° to 118° ± 3°, indicating a hydrophobic surface (Figure 2(b)). Interestingly, there is a sharp decrease in the contact angle after coating the surfaces with DMIMCl and DMIMCl/PDMS composite (Fig. 2(a)). It could happen due to partial filling of the air gap and an increase in surface energy.

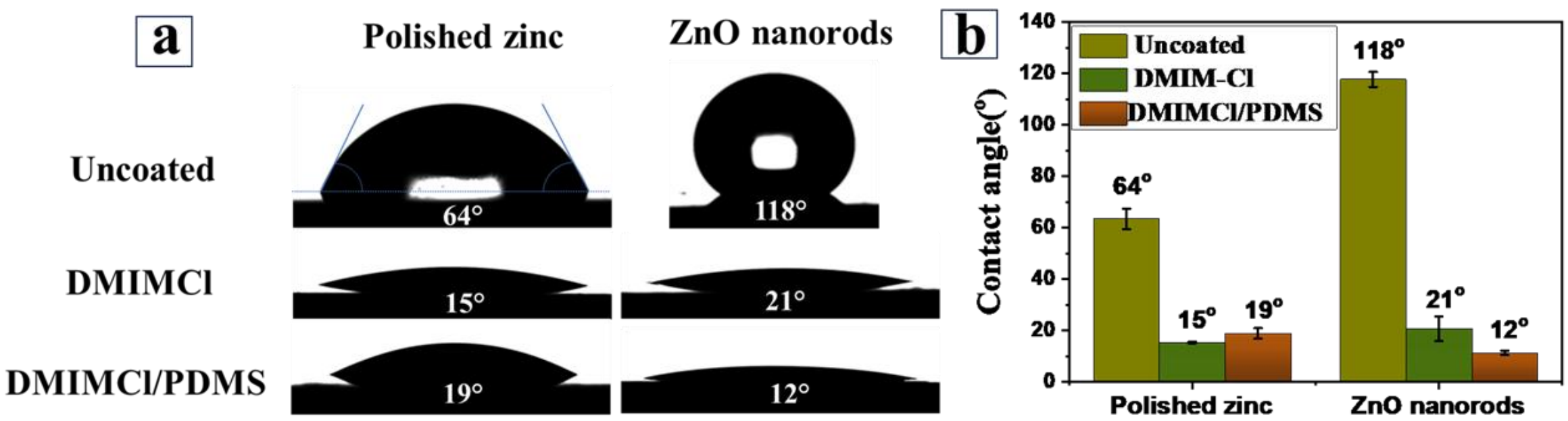

*Figure 2: (a) The optical images of a 2 μL water droplet on Zn substrate (left) and ZnO nanostructured surface (right). (b) Change in the contact angle on these surfaces with coatings of ionic liquid DMIMCl and its mixture with polydimethylsiloxane (PDMS).*

The solid fraction ($f_S$) is defined as the fraction of the projected surface area that is in direct contact with the liquid droplet of a surface. This can be measured by the ImageJ software. This quantity is a critical parameter for determining the wettability and adhesive force of a surface.[25] Here, the grayscale FESEM images were converted into binary white and black maps, with white regions representing the predicted liquid-solid contact area. Figure 3 represents the surface morphology of the zinc substrates before and after chemical modification. When the surfaces are coated with DMIMCl and DMIMCl/PDMS, the surfaces become smoother than the uncoated ones. Due to the capillary effect, the coating material penetrates the grooves and partially fills them, reducing the area fraction of air gaps. This change is directly reflected in the post-coating images (Figure 3(b, c)). The solid area fraction increases from $0.44 \pm 0.02$ for the uncoated surface to $0.54 \pm 0.06$ for DMIMCl-coated surfaces, reaching a maximum of $0.70 \pm 0.02$ for DMIMCl/PDMS-coated surfaces (Figure 3(d)).

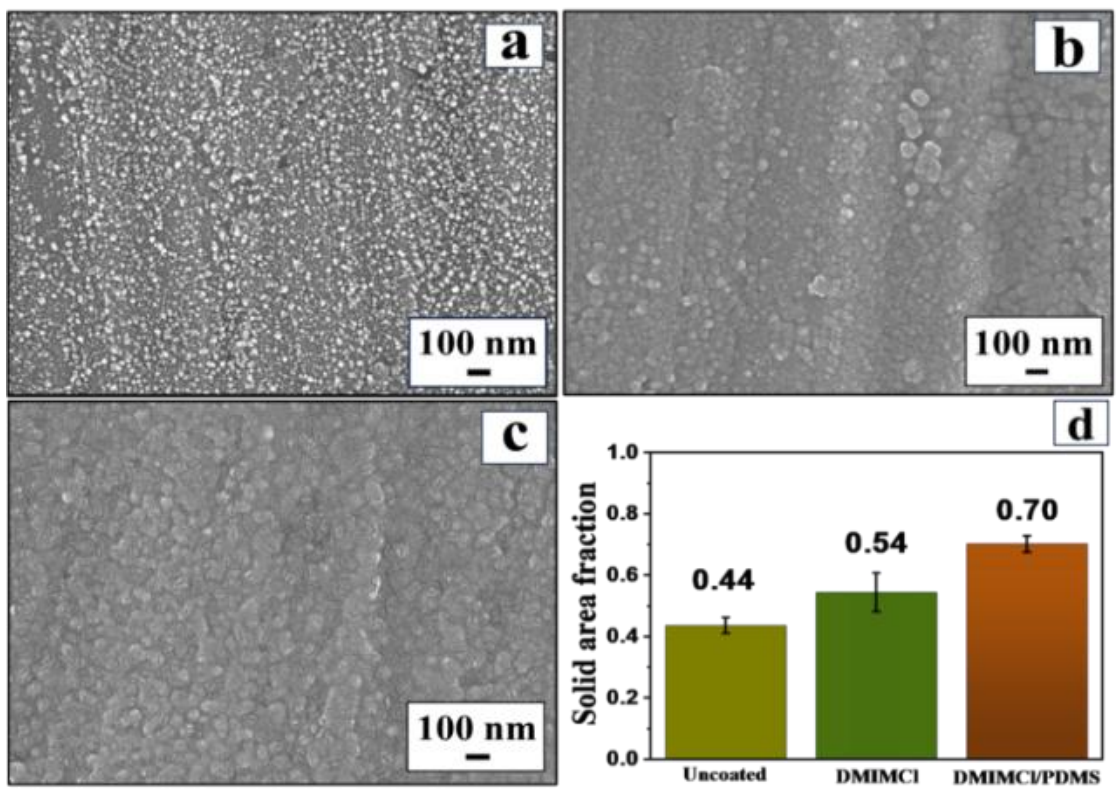

*Figure 3: FESEM images of (a) an uncoated zinc surface and (b) coated with DMIMCl and (c)DMIMCl/PDMS. (d) Solid fraction of zinc surface with different types of coatings.*

The ionic liquid DMIMCl has a cationic head group, which increases the surface energy, allowing the liquid to wet the surface more effectively and reducing the contact angle. Similar behaviour is seen with DMIMCl/PDMS coating as the coating contains 95% DMIMCl, which dominates the surface energy. A similar trend is shown in Figure 4(a), (b), and (c), after observing the surface morphology of ZnO nanorods with or without coating. The solid fraction increases from 0.39 on bare ZnO nanorods to 0.67 after coating with DMIMCl, and the rise in solid fraction up to 0.74 for DMIMCl/PDMS coated surface (Figure 4(d)).

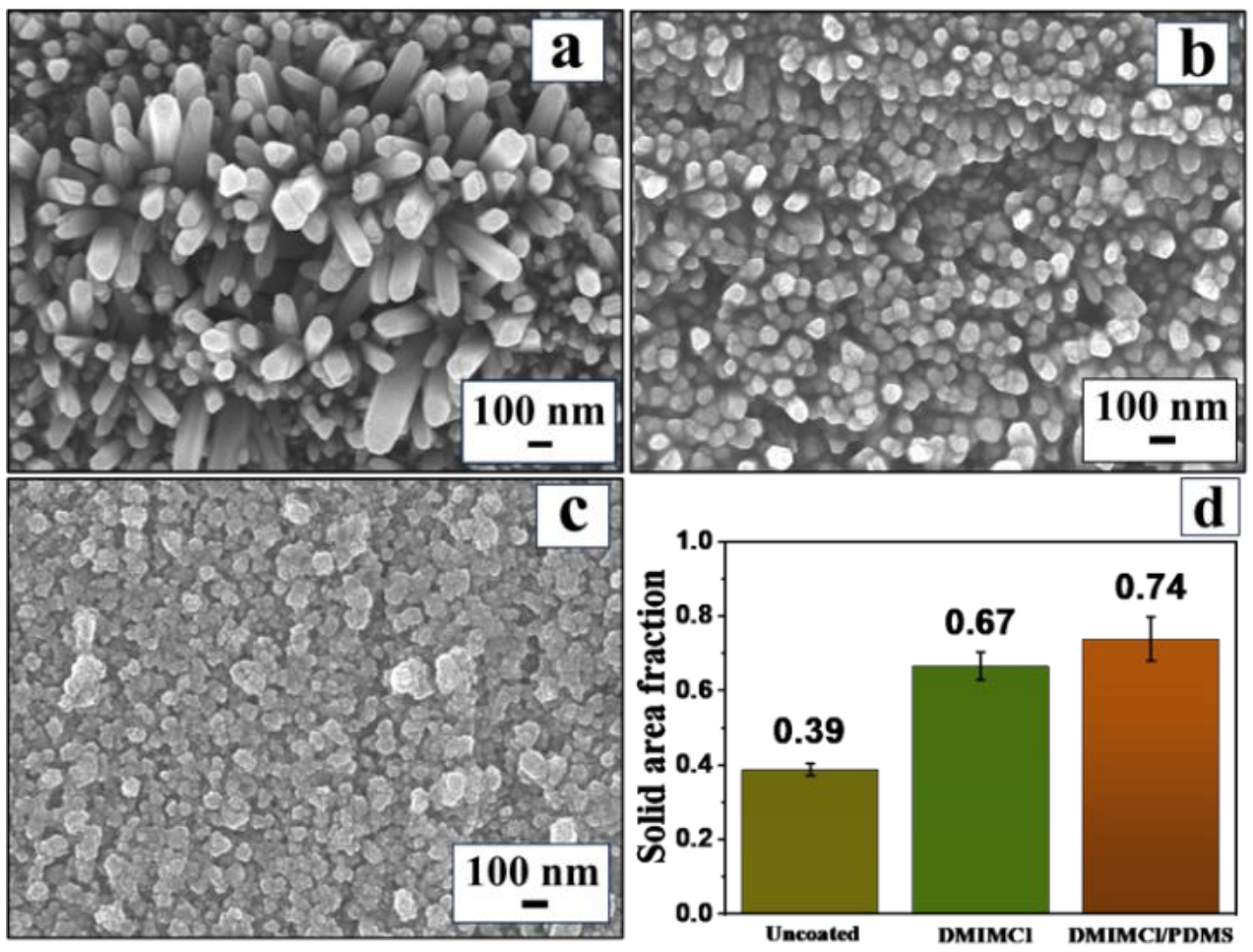

*Figure 4: FESEM images of ZnO nanostructured surface (a) without coating (uncoated) and coated with (b) DMIMCl and (c) DMIMCl/PDMS. (d) Solid fraction of the surfaces with different types of coatings.*

If solid fraction is $f_S$ , then the remaining area $(1-f_S)$ is occupied by trapped air pockets.[27] Therefore, the increase in solid fraction from the zinc substrate to the coated one represents the increase in solid-liquid contact area. As expected for a ZnO nanostructured surface, after coating, and there is a rise in the solid fraction. As a result, the contact angle decreases with increasing solid-liquid contact area and a reduced fraction of trapped air. As explained by Murakami et al., the energy barrier prevents the liquid from entering the grooves, thereby maintaining this state. The upward air pressure successfully opposes the water droplet's downward Laplace pressure on the uncoated nanorods, keeping it there at the top of the structures.[28] The calculated solid surface fraction of the ZnO nanostructured surface was less than 40%, which is consistent with the theoretical prediction of Duez et al. that a reduction in the solid fraction decreases the liquid–solid contact area and promotes the Cassie–Baxter wetting state, thereby enhancing the hydrophobic nature of the surface.[29,30]

Cassie Baxter's stable state of a liquid droplet on a hierarchical surface is observed when $\cos\theta_Y < \frac{f_S-1}{r-f_S}$ , where, $\theta_Y$= contact angle on smooth surface, $f_S$ = solid fraction, and $r$ = roughness factor.[29] Using r = 1.40 ± 0.03 and $f_S$ = 0.39 ± 0.01, right-hand term yields a value of ~ - 0.60, whereas $\cos\theta_Y \sim 0.44$ for $\theta_Y$= 64 °. Therefore, the droplet does not exhibit a stable Cassie-Baxter state. Rather, the air pockets may stay entrapped beneath the water droplet while the liquid droplet partially penetrates the rough surface. This is typical of the metastable Cassie state. The contact angle hysteresis (CAH) was measured to assess the droplet pinning behaviour on the ZnO nanorod surfaces and the value is found to be 100˚. Furthermore, images of water droplets of size 5 µL on the ZnO nanorod surface at tilt angles of 0°, 90°, and 180° are shown in Figure 5. Strong contact-line pinning and high droplet retention are evident in droplets remaining stuck to the surface even in an upside-down position. Hence, high contact angle hysteresis and strong pinning behaviour are consistent with the metastable state, which is a characteristic of the petal effect.[31]

### 3.2 Stability of fabricated surfaces and their adhesion force

Along with morphological stability, it is necessary to assess the firmness of the coatings on a surface for potential applications.[15] To test the stability of our samples, the substrates were immersed in DI water for a period, and the contact angle of a water droplet was measured after drying with $N_2$ gas. The contact angle increases after 16 hours of immersion on the surface coated with DMIMCl (Figure 6(a)). However, the wettability of the DMIMCl/PDMS-coated substrate remains the same throughout the immersion time, showing a nearly constant contact angle. Only 5 wt% PDMS acts as a glue to retain the ionic liquid on the surface. It has a methyl-rich siloxane network and a chemically inert Si-O backbone that does not dissolve in water and remains intact.[32] The surface with ZnO nanorods also shows better stability with the DMIMCl/PDMS coating (Figure 6(b)). Therefore, mixing a few wt% PDMS with a coating material is an effective strategy to enhance the coating's stability.

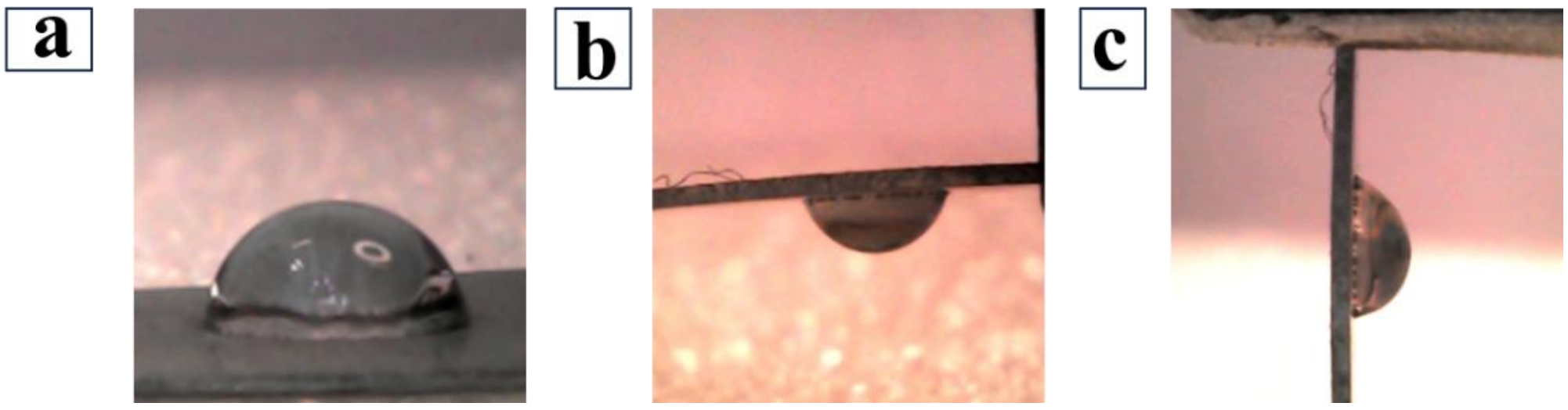

*Figure 5: Optical images of a 5µl water droplet on the nanostructured ZnO surface at (a) horizontal surface, (b)upside-down position and (c) vertical inclination of the surface, demonstrating strong droplet adhesion.*

The decrease in contact angle values due to the coating indicates an increase in the adhesive force between the surface and the water droplet. It is indeed the case, as shown in Figure 6(c). The force on the polished zinc surface without any fabricated nanostructure is measured to be ~563 µN and on DMIMCl-coated zinc is ~743 µN, which increases to ~762 µN for the surface with the DMIMCl/PDMS coating. There is an increase in adhesion force due to the addition of 5 wt% polymer. It is to be noted that the substrate with ZnO nanorods shows a lower force of 332 µN, as air pockets cover a major fraction of the surface. Again, this nanostructured surface shows a higher force when coated with the respective materials. As mentioned earlier, the high-energy ionic liquid enhances the adhesive force, helping water droplets spread on the surface. With the addition of the polymer PDMS, the adhesion force increases for ZnO nanorods, similar to that for the polished zinc surface.

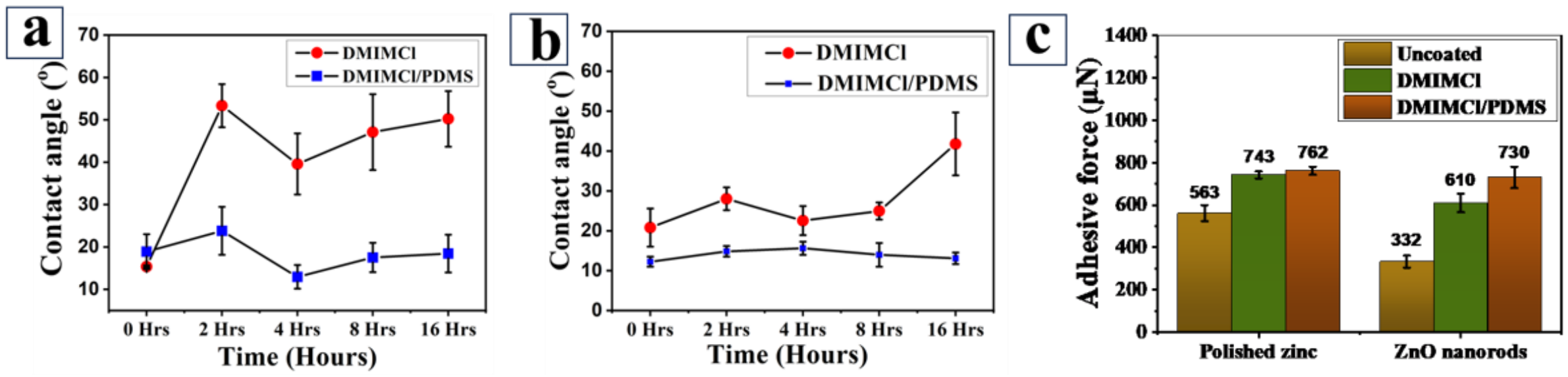

*Figure 6: Water contact angle measurements of DMIMCl and DMIMCl/PDMS coated surfaces of (a) zinc, (b) ZnO nanorods after 2 hours, 4 hours,8 hours and 16 hours of immersion in DI water. (c) (a) Adhesion force measurements for zinc and ZnO nanostructured surfaces with different surface treatments: uncoated, DMIMCl coated, and DMIMCl/PDMS hybrid-coated samples.*

### 3.3 Bacterial efficiency of fabricated surfaces

ZnO nanostructures have attracted considerable attention as functional materials due to their inherent antibacterial properties.[33] In the present study, the antibacterial activities of the fabricated surfaces, with and without coatings of DMIMCl and DMIMCl/PDMS, were evaluated against the Gram-negative bacterium *E. coli* using a spot assay for qualitative analysis and plate counting for quantitative analysis to determine bacterial viability. A spot assay study is performed with different dilution series. Figure 7 shows the results of the spot assay of all the uncoated and coated surfaces. The uncoated samples are used as the control. As observed in Figure 7(a) and (b), there is a significant reduction in colony formation for coated samples, whereas the uncoated control sample exhibited substantial bacterial growth. As observed in Figure 7, there is a statistically significant decrease in bacterial growth even on the uncoated surface due to the inherent antibacterial activity of the zinc metal. The higher dilution series shows that both the coated and uncoated surfaces reduce the number of colonies, in some cases, even eliminate them. It can be inferred that the nanostructured surface inhibits bacterial growth compared with the polished zinc surface. Among the coated samples, the nanostructured sample shows greater reduction in the 4$^{th}$ and 5$^{th}$ dilution series. The ionic liquid DMIMCl greatly reduces bacterial colonies on surfaces by rendering them highly antibacterial. As discussed in the previous section, the difference in their stability, however, makes the DMIMCl/PDMS coating a better choice.

The antibacterial properties of the surfaces were further quantified using the colony-counting method. Figure 8(a) of the polished surface shows many bacterial colonies appearing as white spots, which are reduced on DMIMCl and DMIMCl/PDMS-coated surfaces. Interestingly, the uncoated nanostructured surface shows a considerable reduction in white spots compared to the pristine polished surface, suggesting the effectiveness of the nanostructured surface in controlling bacterial growth. As evident in Figure 8(b), the coated nanostructured surfaces exhibit further reduction. The quantitative bar graph of Figure 8(c) indicates that the DMIMCl-coated surface is slightly more active compared to the DMIMCl/PDMS one. The number of colonies decrease by 61% on the DMIMCl-coated polished zinc substrate and by 71% on the DMIMCl-coated ZnO nanostructured substrate. As discussed earlier, the presence of PDMS provides better stability to the coating, indicating it is a better coating material for practical purposes.

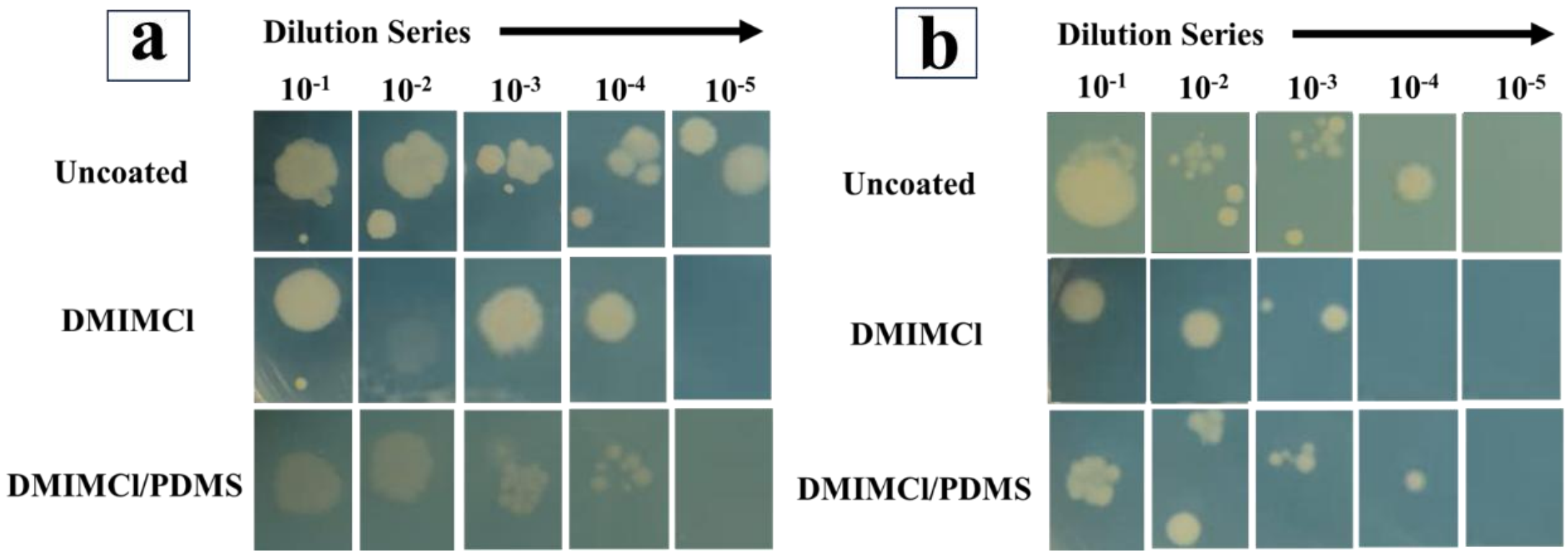

*Figure 7: Spot assay of bacterial growth after 6 hours of incubation on (a) polished surface, and (b)ZnO-nanorods decorated surface without (uncoated) and coated with ionic liquid DMIMCl and DMIMCl/PDMS.*

Because of the hydrophilic nature of a surface, bacteria easily adhere to it, as the outer surface of bacterial cells is also hydrophilic, primarily due to the presence of lipopolysaccharides. While the bacteria adhered to the surface, the nanostructured surface, with dimensions of a few nanometres, can pierce the bacterial cell, E. coli, which has a size of a few µm. This mechanism is well established in the literature.[34, 35,]

The uniqueness of the present study is the environmentally friendly HWT approach for developing such a surface. On the other hand, various ionic liquids have been investigated to quantify their antibacterial activities in solution but not as a coating materials.[36] The ionic liquids are amphiphilic in nature with a cationic head group, especially for imidazolium-based

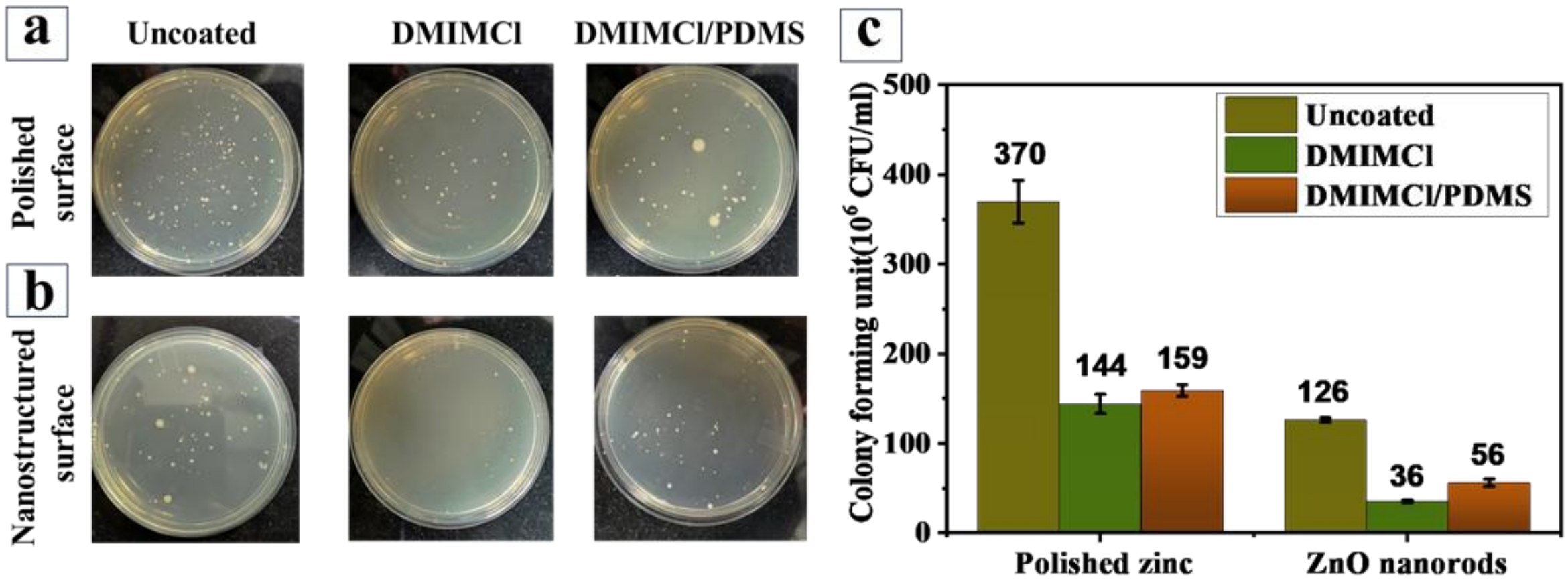

*Figure 8: Bacterial colonies (white spots) developed on uncoated, DMIMCl coated and DMIMCl/PDMS coated substrates of (a) polished surface and (b) nanostructured surface. (c) The quantity of Colony Forming Units (CFU) after 6 hours of incubation on respective surfaces*

ionic liquids, as is used in the present study, and a hydrocarbon chain. This head group electrostatically interacts with the negatively charged lipids of the bacterial cell membrane, thereby facilitating the insertion of their hydrocarbon chains into the hydrophobic core of the membrane.[36] This insertion destabilizes the physiologically relevant membrane of the bacteria, causing them to die. In the present study, this property of the ionic liquid has been utilized to develop an effective antimicrobial surface by coating the ZnO surface.

It may be noted that ZnO is well-suited for such applications due to its intrinsic antimicrobial properties. A metal oxide nanostructure produces reactive oxygen species (ROS), including hydroxyl radicals, superoxide anions, and hydrogen peroxide, that chemically interfere with the physiological functions of the bacterial cell membrane and intracellular macromolecules.[37,38]

The Wenzel equation [39], $cos\,\theta_W = r.cos\,\theta_Y$, dictates that introducing roughness on an ideal surface can only amplify the wettability of the ideal surface. In our case, the ideal Zn surface is intrinsically hydrophilic ($\theta_Y$ = 64°, cos$\theta_Y$= 0.44), and after forming nanostructures on zinc surface having roughness factor r is ≥ 1. Wenzel equation predicts that this should make the surface more hydrophilic ($cos\,\theta_W \geq 0.44$ and $\theta_W \leq 64°$). The pronounced hydrophobic response observed in this study is therefore incompatible with homogeneous Wenzel wetting and instead points to a composite, air-trapping interface described by the Cassie-Baxter model.[40, 41] A careful analysis suggests that the state is actually a metastable one. These results indicate that the trapped air within the nanorod array is the main driving force behind the wettability transition, consistent with the roughness-mediated wetting framework established for nanostructured surfaces.[31] The coating of DMIMCl reversed this trend, lowering the contact angle and increasing the adhesive force from 332 μN (bare nanorods) to 730 μN, reflecting the high surface energy of the imidazolium head group. Incorporation of a fraction of PDMS (5 wt%) preserves this chemical effect while enhancing the coating's durability under aqueous immersion, consistent with PDMS's chemically inert siloxane backbone.[42]

Testing our synthesized surfaces for antibacterial properties against *E. coli* showed a clear trend of efficacy: polished Zn < ZnO nanorods < ILcoated ZnO nanorods. This supports a triple-mechanism model in which nanotopography-induced mechanical stress, ZnO-mediated reactive oxygen species, and ILs' membrane-disruptive abilities act synergistically to enhance bacterial killing. This effective advancement in the physicochemical design was achieved using green synthesis while limiting toxicity associated with the chemical synthesis approach.

# 4. Conclusions

In this study, ZnO nanorod arrays were fabricated on zinc substrates using an environmentally friendly, toxic chemical free hot-water treatment (HWT) method, and their surface wettability, adhesion, and antibacterial performance were systematically tuned by coating them with DMIMCl alone or with a DMIMCl-PDMS composite. The synthesis of ZnO nanostructures altered the surface characteristics from hydrophilic to hydrophobic. Coating with DMIMCl reversed the wettability trend and substantially increased surface adhesion, raising the adhesive force, while incorporation of a small PDMS fraction (5 wt%) preserved this chemical effect and improved coating durability under aqueous immersion. Antibacterial testing against E. coli demonstrated a clear, progressive enhancement in bactericidal efficacy across the sample series, with colony-forming units dropping from uncoated polished zinc to uncoated ZnO nanorods to the least number of colonies on the coated nanorod surfaces. This trend supports a triple-mechanism model in which topography-induced mechanical stress, ZnO-mediated reactive oxygen species generation, and the membrane-disruptive action of the ionic liquid act together to kill bacteria more effectively than any single mechanism alone.

From an application point of view, the combination of green synthesis, tunable wettability, enhanced adhesion, and low-temperature processing positions the present approach as a candidate for antibacterial and anti-fouling surfaces. These surfaces are best suited for healthcare (high-touch and implantable surfaces), food storage, and marine anti-fouling. However, translating these results into any of these settings will require closing several gaps left open by the current study. First, antibacterial evaluation was limited to a single Gram-negative species, given the documented differences in mechanical and membrane susceptibilities between Gram-negative and Gram-positive bacteria. Extending efficacy tests to Gram-positive strains (e.g., *S. aureus*) is necessary before broader efficacy claims can be made. The coating stability was assessed only up to a 16-hour immersion window, but for real-life applications, such as healthcare and marine applications, long-term immersion testing and saline stability testing are required. Finally, as imidazolium ionic liquids are known to interact with lipid membranes [43,44,45], the biocompatibility and cytotoxicity of this coating toward mammalian cells should be established before healthcare or food-based applications. Addressing these points could lead to a more practically deployable, sustainable/green synthesized antibacterial surface technology built on this platform.